\documentclass[aps,prd,twocolumn,10pt,superscriptaddress,nofootinbib,nobibnotes,longbibliography]{revtex4-1}

\usepackage{amssymb}
\usepackage{graphicx}
\usepackage{amsmath}
\usepackage{hyperref}
\usepackage{subfigure}
\usepackage{multirow}
\usepackage{setspace}
\usepackage{verbatim}
\usepackage{float}
\usepackage{color}
\usepackage{ulem}
\usepackage[utf8]{inputenc}
\usepackage[table,xcdraw]{xcolor}
\usepackage{makecell}
\usepackage{url}
\usepackage{bm}
\usepackage{placeins}
\usepackage{slashed}

\newcommand{\md}{\mathrm{d}}

\begin{document}

\title{Thermodynamics of Kerr-Bertotti-Robinson black hole}

\author{Li Hu}
\email{huli21@mails.ucas.ac.cn}
\affiliation{School of Fundamental Physics and Mathematical Sciences, Hangzhou Institute for Advanced Study (HIAS), University of Chinese Academy of Sciences (UCAS), Hangzhou 310024, China}
\affiliation{Institute of Theoretical Physics, Chinese Academy of Sciences (CAS), Beijing 100190, China}
\affiliation{University of Chinese Academy of Sciences (UCAS), Beijing 100049, China}

\author{Rong-Gen Cai}
\email{caironggen@nbu.edu.cn}
\affiliation{Institute of Fundamental Physics and Quantum Technology, \& School of Physical Science and Technology, Ningbo University, Ningbo, 315211, China}

\author{Shao-Jiang Wang}
\email{schwang@itp.ac.cn (Corresponding author)}
\affiliation{Institute of Theoretical Physics, Chinese Academy of Sciences (CAS), Beijing 100190, China}
\affiliation{Asia Pacific Center for Theoretical Physics (APCTP), Pohang 37673, Korea}

\begin{abstract}
We investigate the thermodynamic properties of the Kerr-Bertotti-Robinson black hole, an exact Petrov type-D solution of Einstein-Maxwell theory describing a rotating black hole immersed in an external electromagnetic field. While the conserved angular momentum and electric charge can be computed straightforwardly, the conserved mass cannot be obtained through standard integrability methods due to the nontrivial asymptotically uniform external electromagnetic field. To overcome this difficulty, we adopt the Christodoulou-Ruffini mass relation as a thermodynamic definition of the conserved mass and identify the associated generator, thereby fixing the ambiguity in defining this conserved mass and constructing the thermodynamic potentials. These thermodynamic quantities naturally satisfy the first law of black-hole thermodynamics as well as the Smarr formula.
\end{abstract}
\maketitle

\section{Introduction} 

Black holes play a central role in both theoretical physics and astronomy. Recent advances, including the direct detection of gravitational waves from binary black hole mergers by LIGO and Virgo~\cite{LIGOScientific:2016aoc} and the direct imaging of black-hole shadows by the Event Horizon Telescope~\cite{EventHorizonTelescope:2019dse}, have provided compelling evidence for their astrophysical existence. These developments further motivate the study of realistic black-hole environments beyond idealized vacuum solutions~\cite{Hu:2023oiu, Hu:2024hzu, Vicente:2025gsg, Lyu:2025lue}. 

Astrophysical black holes, particularly those residing in galactic centers~\cite{Richstone:1998ky, Kormendy:2013dxa}, are typically surrounded by accretion disks that can sustain strong magnetic fields~\cite{Eatough:2013nva, You:2023dax}. Therefore, constructing exact solutions that incorporate external electromagnetic fields has always been a focus of theoretical research. Early progress in this direction was made by Ernst and Wild~\cite{Ernst:1976bsr, Ernst:1976mzr}, who obtained the Kerr-Newman-Melvin spacetime describing a rotating charged black hole immersed in a magnetic universe. However, this solution exhibits several limitations, such as the presence of an ergoregion extending to infinity~\cite{Gibbons:2013yq} and the algebraic type-I structure~\cite{Pravda:2005uv}, which restrict its applicability in astrophysical contexts.

More recently, Podolsk\'{y} and Ovcharenko have successfully constructed a new exact solution with magnetic field, dubbed the Kerr-Bertotti-Robinson (Kerr-BR) spacetime~\cite{Ovcharenko:2025cpm, Podolsky:2025tle}. This geometry describes a rotating black hole embedded in an external electromagnetic field with improved algebraic structure (Petrov type D) and more tractable properties. As a result, it has attracted considerable attention in recent studies, including investigations of energy extraction via magnetic Penrose processes~\cite{Zeng:2025olq, Mirkhaydarov:2026fyn}, analyses of black-hole shadows~\cite{Wang:2025vsx, Ali:2025beh, Liu:2025wwq}, extended black hole solutions~\cite{Ahmed:2025ril,Ovcharenko:2026byw,Al-Badawi:2026whl,Astorino:2025lih,Barrientos:2026shy}, and the influence of external magnetic fields on gravitational waves~\cite{Li:2025rtf, Xamidov:2026kqs, Yuan:2026knu}, to name just a few. For other studies, see Refs.~\cite{Siahaan:2025ngu, Zeng:2025tji, Wang:2025bjf, Vachher:2025jsq, Zhang:2025ole, Ortaggio:2025sip, Gray:2025lwy, Andersson:2025bhq, Astorino:2026okd, Barrientos:2026kdl, Wang:2026czl, Mustafa:2026gly, Siahaan:2026tuf, Lu:2026kcm}.

Despite these developments, the thermodynamic description of the Kerr-BR black hole remains largely unexplored. In this paper, we present the conserved angular momentum $J$, electric charge $Q$, and the mass $M$ for the first time. Owing to the nontrivial asymptotic structure, the identification of the appropriate generator associated with the conserved mass is not straightforward. Consequently, the covariant phase space formalism does not directly yield an integrable mass through standard procedures~\cite{Gibbons:2013dna, Astorino:2016hls, Astorino:2016xiy, Astorino:2016ybm, Astorino:2025lih, Gao:2023luj}. To address this issue, we adopt the Christodoulou-Ruffini mass formula as a thermodynamic definition of the conserved mass~\cite{Astorino:2016hls, Astorino:2016ybm}. This formula, originally derived from the limit of electric Penrose processes~\cite{Christodoulou:1971pcn, Hu:2025bbc}, provides a functional dependence $M=M(S,J,Q)$ that implicitly encodes thermodynamic consistency.

Based on this definition, we determine the generator associated with the conserved mass in terms of the black hole mass, spin, and electromagnetic field strength, which turns out to be a nontrivial combination. We then construct the redefined thermodynamic potentials and show that they coincide with those derived directly from the Christodoulou-Ruffini relation. As a consequence, all explicit dependence on the external magnetic field in the first law of black hole thermodynamics and Smarr formula is absorbed into the thermodynamic quantities, so both relations take their standard forms without additional $\mu \delta B$ or $\mu B$ terms, in agreement with previous results for magnetized black holes~\cite{Astorino:2016hls}. This establishes a consistent thermodynamic description of the Kerr-BR black hole in the presence of external magnetic fields.


\section{Kerr-BR black hole}

The Kerr-BR metric~\cite{Ovcharenko:2025cpm, Podolsky:2025tle} looks explicitly like
\begin{align}
    \md s^2&=\frac{1}{\Sigma^2}\bigg[-\frac{\beta}{\rho^2}\big(\md t-a\sin^2\theta\,\md\phi\big)^2+\frac{\rho^2}{\beta}\md r^2+\frac{\rho^2}{P}\md\theta^2\nonumber\\
    &\qquad\qquad+\frac{P}{\rho^2}\sin^2\theta\,\big(a\md t-(r^2+a^2)\md\phi\big)^2\bigg]\,,
\end{align}
where
\begin{align}
    \rho^2&=r^2+a^2\,\cos^2\theta\,,\\
    P&=1+B^2\Big(m^2\frac{I_2}{I_1^2}-a^2\Big)\cos^2\theta\,,\\
    \beta&=\big(1+B^2r^2\big)\Delta\,,\\
    \Sigma^2&=\big(1+B^2r^2\big)-B^2\Delta\cos^2\theta\,,\\
    \Delta&=\Big(1-B^2m^2\frac{I_2}{I_1^2}\Big)r^2-2m\frac{I_2}{I_1}r+a^2\,,\\
    I_1&=1-\frac{1}{2}B^2a^2\,,\\
    I_2&=1-B^2a^2\,.
\end{align}
The three basic parameters $m$, $a$, and $B$ stand for the black-hole mass, spin, and electromagnetic field strength, respectively.

The corresponding gauge potential of this asymptotically uniform electromagnetic field can be described as
\begin{align}
    A_\mu \md x^\mu&=\frac{\mathrm{e}^{\mathrm{i}\,\gamma}}{2B}\Big[\Sigma,_{\,r}\frac{a\md t-(r^2+a^2)\md\phi}{r+ia\cos\theta}+(\Sigma-1)\md\phi\nonumber\\
    &\qquad\qquad+\frac{i\Sigma,_{\,\theta}}{\sin\theta}\frac{\md t-a\sin^2\theta\md\phi}{r+ia\cos\theta}\Big]\,,
\end{align}
where the parameter $\gamma$ represents the duality rotation. In case of $m=0$, $\gamma = 0$ yields a pure magnetic field, and $\gamma = \pi/2$ yields a pure electric field, though $\gamma$ itself does not influence the strength parameter $B$~\cite{Siahaan:2025ngu}. Furthermore,  the physical gauge potential is defined as the real counterpart $A_\mu^\mathrm{real} \equiv 2\,\mathrm{Re}\,A_\mu$, and the associated electromagnetic field $F_{\mu\nu}=\partial_\mu \mathrm{A}_\nu^\mathrm{real}-\partial_\nu \mathrm{A}_\mu^\mathrm{real}$ satisfies the Einstein-Maxwell equations, as expected.

Before discussing the thermodynamic properties, several modifications must be made to the Kerr-BR metric and the associated gauge potential. First, we choose an appropriate normalization of the axial Killing vector $\partial_\phi$ in order to remove possible conical defects along the symmetry axis. The redefined angular coordinate $\varphi$ is required to have the standard periodicity $2\pi$. Following Ref.~\cite{Podolsky:2025tle}, this new angle is introduced as $\varphi=\phi \,P_0$, where the normalization factor $P_0$ is determined by comparing the proper circumference and radius of small circles around the poles $\theta=0$ and $\theta=\pi$,
\begin{align}
    P_0=1+B^2\Big(m^2\frac{I_2}{I_1^2}-a^2\Big)\,.
\end{align}

The second modification concerns the regularity of the gauge potential on the rotation axis. We require that the real component of the azimuthal gauge potential $A^\mathrm{real}_\varphi$ vanish at both the north and south poles, ensuring that the gauge potential is regular along the axis. This condition can be satisfied by exploiting the gauge freedom and shifting $A^\mathrm{real}_\varphi$ by a real constant $A_0$, making use of the inherent gauge freedom. A subtle point is that avoiding the Dirac string restricts the duality rotation parameter: one must set $\gamma=0$, corresponding to a purely magnetic external field. The required constant shift is therefore
\begin{align}
    A_0&=\frac{\sqrt{I_2}-1}{P_0B}\,.
\end{align}

One may naturally ask whether the magnetic charge $U$ vanishes only after imposing $\gamma=0$. We have explicitly verified that, for arbitrary values of the duality angle $\gamma$, the magnetic charge vanishes identically,
\begin{align}
    U=\frac{1}{8\pi}\int_{\partial C}F_{\mu\nu}\,\mathrm{d}x^\mu\wedge\mathrm{d}x^\nu=\frac{1}{4\pi}\int_{\partial C}F_{\theta\varphi}\,\mathrm{d}\theta\,\mathrm{d}\varphi=0,
\end{align}
where $C$ denotes a Cauchy surface. Therefore, the absence of magnetic charge is an intrinsic property of the Kerr-BR solution rather than a consequence of imposing the regularity condition.

After implementing these two modifications, the metric and the gauge potential take the form
\begin{align}
    \md s^2&=\frac{1}{\Sigma^2}\bigg[-\frac{\beta}{\rho^2}\big(\md t-\frac{a\sin^2\theta}{P_0}\,\md\varphi\big)^2+\frac{\rho^2}{\beta}\md r^2+\frac{\rho^2}{P}\md\theta^2\nonumber\\
    &\qquad\qquad+\frac{P}{\rho^2}\sin^2\theta\,\big(a\md t-\frac{r^2+a^2}{P_0}\md\varphi\big)^2\bigg]\,,
\end{align}
and
\begin{align}
    A_\mu^\mathrm{real}\md x^\mu&=\frac{1}{B}\Big[\md t\frac{1}{\rho^2}\big(ar\,\Sigma,_{\,r}+\frac{a\,\Sigma,_{\,\theta}}{\tan\theta}\big)+\frac{\Sigma-\sqrt{I_2}}{P_0}\md\varphi\nonumber\\
    &-\md\varphi\frac{1}{\rho^2P_0}\Big(r(r^2+a^2)\Sigma,_{\,r}+a^2\sin\theta\cos\theta\,\Sigma,_{\,\theta}\Big)\Big]\,.
\end{align}
With these regularity conditions imposed, we can now proceed to compute the thermodynamic parameters defined at the outer horizon.

\section{Thermodynamical quantities}

We now determine the thermodynamic quantities associated with the outer event horizon. 
The location of the outer horizon is given by the largest root of the function $\Delta$, that is~\cite{Podolsky:2025tle},
\begin{align}
    r_+=\frac{mI_2+\sqrt{m^2I_2-a^2I_1^2}}{I_1^2-B^2m^2I_2}I_1\,,
\end{align}
showing that the position of the outer horizon is modified by the presence of the external electromagnetic field $B$.

The angular velocity of the outer horizon is obtained from the standard relation,
\begin{align}
    \Omega_{H}=-\left.\frac{g_{t\varphi}}{g_{\varphi\varphi}}\right|_{r=r_+}=\frac{aP_0}{r_+^2+a^2}\,.
\end{align}

The Hawking temperature $T_H$ and the Bekenstein-Hawking entropy $S$ follow from the surface gravity $\kappa$ and the horizon area $\mathcal{A}$, respectively~\cite{Bekenstein:1973ur, Bardeen:1973gs, Siahaan:2025ngu},
\begin{align}
    T_H&=\frac{\kappa}{2\pi}=\frac{1}{2\pi}\frac{1+B^2r_+^2}{r_+^2+a^2}\Big(m\frac{I_2}{I_1}-\frac{a^2}{r_+}\Big)\,,\\
    S&=\frac{\mathcal{A}}{4}=\frac{\pi}{P_0}\frac{r_+^2+a^2}{1+B^2r_+^2}\,.
\end{align}

The Coulomb electrostatic potential is defined with respect to the Killing generator of the horizon $\chi=\partial_t+\Omega_H\partial_\varphi$. Evaluating the gauge potential along this vector at the outer horizon gives rise to the electrostatic potential as
\begin{align}
    \Phi_H&=-A^\mathrm{real}_\mu\chi^\mu|_{r=r_+}\nonumber\\
    &=\frac{-Ba\left[4-4B^2(m^2+a^2)+B^4a^2(4m^2+a^2)\right]}{B^4m^3a^2I_2+mI_1^2(2-3B^2a^2)+2I_1^2\sqrt{m^2I_2-a^2I_1^2}}\nonumber\\
    &\quad\times\frac{m(1-B^2a^2)+\sqrt{m^2I_2-a^2I_1^2}}{4\sqrt{I_2}}\,.
\end{align}
It is worth noting that the electrostatic potential is independent of the polar angle $\theta$, as expected for a regular Killing horizon.

\section{Conserved charges}

The Kerr-BR metric is derived from the standard Einstein-Maxwell action,
\begin{align}
    S=\frac{1}{16\pi}\int\md^4x\sqrt{-g}\,(R-F_{\mu\nu}F^{\mu\nu})\,.
\end{align}
The generalized Killing equations for fields $g_{\mu\nu}$ and $A_\mu$ can be expressed as
\begin{align}
    \mathcal{L}_{\xi}g_{\mu\nu}&=0\,,\\
     \mathcal{L}_{\xi}A_{\mu}+\partial_{\mu}\lambda&=0\,,
\end{align}
where the gauge parameters $\xi=\xi^\mu\partial_\mu$ and $\lambda$ are a Killing vector field and a real constant, respectively.


According to covariant phase space methods (a formalism pioneered by Iyer-Wald~\cite{Iyer:1994ys} and further developed by Barnich-Brandt~\cite{Barnich:2001jy, Barnich:2003xg} to address issues of asymptotic integrability), the total conserved charge $\mathcal{Q}$ depends on the choice of gauge parameters $(\xi,\lambda)$ and is defined as
\begin{align}
    \mathcal{Q}_{(\xi,\lambda)}[g,A;\bar{g},\bar{A}]=\int_{\partial C}\int_{\bar{g}}^g\int_{\bar{A}}^A  k_{(\xi,\lambda)}[\delta g',\delta A';g',A']\,,
\end{align}
where $C$ is a Cauchy surface, $\bar{g}_{\mu\nu}$ and $\bar{A}_\mu$ are a set of background solutions, and the explicit form of the surface charge 2-forms $k_{(\xi,\lambda)}$ is adopted from Refs.~\cite{Barnich:2001jy, Compere:2009dp},
\begin{align}
    k_{(\xi,\lambda)}[\delta g,\delta A;g,A]=k_{\xi}^{grav}+k_{(\xi,\lambda)}^F\,.
\end{align}
Here,
\begin{subequations}\label{eq:surCharge}
\begin{align}
    k_{\xi}^{grav}&=\frac{1}{8\pi}(\md^2x)_{\mu\nu}\Big\{\xi^\nu\nabla^\mu h-\xi^\nu\nabla_\rho h^{\mu\rho}+\xi_\rho\nabla^\nu h^{\mu\rho}\nonumber\\
    &\quad+\frac{1}{2}h\nabla^\nu\xi^\mu-h^{\rho\nu}\nabla_\rho\xi^\mu\Big\}\,,\\
    k_{(\xi,\lambda)}^F&=\frac{1}{4\pi}(\md^2x)_{\mu\nu}\bigg\{-F^{\mu\nu}\delta A_\rho\xi^\rho-2\xi^\mu F^{\nu\rho}\delta A_\rho\nonumber\\
    &\quad+\left[2h^{\mu\rho}F_\rho^{\;\nu}-\delta F^{\mu\nu}-\frac{1}{2}hF^{\mu\nu}\right]\left(A_\sigma\xi^\sigma+\lambda\right)\bigg\}\,,
\end{align}
\end{subequations}
where we define 
\begin{align}
    (\md^2x)_{\mu\nu}&=\frac{1}{4}\epsilon_{\mu\nu\rho\sigma}\,\md x^\rho\wedge\md x^\sigma\,,\qquad\epsilon_{tr\theta\varphi}=1\,,\nonumber\\
    h_{\mu\nu}&=\delta g_{\mu\nu}\,,\qquad h^{\mu\nu}=g^{\mu\rho}g^{\nu\sigma}h_{\rho\sigma}\,,\qquad h=g^{\mu\nu}h_{\mu\nu}\,,\nonumber\\
    \delta F_{\mu\nu}&=\partial_\mu\delta A_\nu-\partial_\nu\delta A_\mu\,,\qquad\delta F^{\mu\nu}=g^{\mu\rho}g^{\nu\sigma}\delta F_{\rho\sigma}\,.\nonumber
\end{align}

Since the integral on the boundary surface $\partial C$ is performed along a path of solutions, the definition is only meaningful if the integrability condition is satisfied~\cite{Compere:2006my},
\begin{align}
    \int_{\partial C}\left(\delta_1 k_{(\xi,\lambda)}[\delta_2 g,\delta_2 A;g,A]-\delta_2 k_{(\xi,\lambda)}[\delta_1 g,\delta_1 A;g,A]\right)=0\,.
\end{align}

The conserved angular momentum $J$ and electric charge $Q$ can be obtained straightforwardly by choosing the gauge parameters $(-\partial_\varphi,0)$ and $(0,-1)$, respectively. In these cases, all boundary terms in Eq.~\eqref{eq:surCharge} vanish identically, rendering the infinitesimal charges $\delta Q$ and $\delta J$ integrable. Their integration yields
\begin{align}
    J&=\mathcal{Q}_{(-\partial_\varphi,0)}=\frac{2ma(2-B^2a^2)^3}{(1-B^2a^2)\big[4+B^4a^4+4B^2(m^2-a^2)\big]^2}\,,\\
    Q&=\mathcal{Q}_{(0,-1)}=\frac{2Bma(-2+B^2a^2)}{\sqrt{1-B^2a^2}\big[4+B^4a^4+4B^2(m^2-a^2)\big]}\,.
\end{align}
In contrast, obtaining the conserved total energy is considerably more subtle, as the Kerr-BR spacetime is not asymptotically flat, making the appropriate generator for the thermodynamic mass obscure. In general, it is not as simple as $(\partial_t,0)$, for which the infinitesimal conserved charge takes the form
\begin{align}\label{eq:slash_mass}
\slashed{\delta}\mathcal{Q}_{(\partial_t,0)}=\int_{\partial C}k_{(\partial_t,0)}=c_m\delta m+c_a\delta a+c_B\delta B\,,
\end{align}
where
\begin{align}
    c_m&=\frac{4I_1\left(4-4B^2a^2-8B^2m^2+8B^4m^2a^2+B^4a^4\right)}{I_2\big[4+B^4a^4+4B^2(m^2-a^2)\big]^2}\,,\\
    c_a&=\frac{2B^2ma}{I_2\big[4+B^4a^4+4B^2(m^2-a^2)\big]^2}\times\big(16-8B^2m^2\nonumber\\
    &\quad-20B^2a^2+12B^4m^2a^2+8B^4a^4-B^6a^6\big)\,,\\
    c_B&=\frac{4Bm}{I_2\big[4+B^4a^4+4B^2(m^2-a^2)\big]^2}\nonumber\\
    &\hspace{-4mm}\times\big[12(a^2-m^2)+B^2a^2(14m^2-16a^2+7B^2a^4-B^4a^6)\big]\,,
\end{align}
and the notation $\slashed{\delta}$ emphasizes that Eq.~\eqref{eq:slash_mass} does not satisfy the integrability condition.

Given that we have three independent generators, a natural approach is to consider a linear combination, such as $\alpha(\partial_t+\Omega_{int}\partial_\varphi,\Phi_{int})$, to characterize the conserved energy
\begin{align}
    \delta M=\alpha\left(\slashed{\delta}\mathcal{Q}_{(\partial_t,0)}-\Omega_{int}\delta J-\Phi_{int}\delta Q\right)\,,
\end{align}
where $\alpha$, $\Omega_{int}$, and $\Phi_{int}$ are undetermined parameters. Expanding these variations gives the following three equations:
\begin{subequations}\label{eq:int_condi}
\begin{align}
    \partial_m M&=\alpha\left(c_m-\Omega_{int}\partial_m J-\Phi_{int}\partial_m Q\right)\,,\label{eq:int_condi_m}\\
    \partial_a M&=\alpha\left(c_a-\Omega_{int}\partial_a J-\Phi_{int}\partial_a Q\right)\,,\label{eq:int_condi_a}\\
    \partial_B M&=\alpha\left(c_B-\Omega_{int}\partial_B J-\Phi_{int}\partial_B Q\right)\,.\label{eq:int_condi_B}
\end{align}
\end{subequations}
There are three equations, but with four unknowns, indicating that the conserved mass cannot be uniquely determined from integrability conditions alone. Fortunately, previous studies indicate that enforcing thermodynamic consistency leads to a mass that coincides with the well-known Christodoulou-Ruffini mass relation~\cite{Astorino:2016hls, Astorino:2016xiy, Astorino:2016ybm, Astorino:2025lih},
\begin{align}
    M^2(S,J,Q)=\frac{S}{4\pi}+\frac{Q^2}{2}+\frac{\pi(Q^4+4J^2)}{4S}\,.
\end{align}
Since in Einstein-Maxwell gravity, the intrinsic parameters of a stationary black hole are given by its entropy $S$, angular momentum $J$, and electric charge $Q$, it is natural to adopt this relation as the definition of the conserved mass. In what follows, we determine the corresponding generator of this conserved mass, and the resulting thermodynamic quantities naturally satisfy the first law of black hole thermodynamics and the Smarr formula.

The resulting Christodoulou-Ruffini mass takes the form
\begin{align}
    M(m,a,B)&=2m(2-B^2a^2)\nonumber\\
    &\hspace{-15mm}\times\sqrt{\frac{\left(4+B^6a^6-3B^4a^4+4B^4m^2a^2\right)}{(1-B^2a^2)\big[4+B^4a^4+4B^2(m^2-a^2)\big]^3}}\,.
\end{align}
This expression possesses several desirable properties. In particular, it reduces to the conserved mass of the Schwarzschild-Bertotti-Robinson spacetime in the nonrotating limit $a\rightarrow 0$~\cite{Astorino:2025lih}, and it reproduces the standard Kerr mass when the external magnetic field is switched off $B\rightarrow 0$. This provides further support for identifying the above expression as the physical conserved mass.

Besides, substituting this mass into Eq.~\eqref{eq:int_condi} immediately determines the three previously undetermined parameters in a linear combination of generators,
\begin{align}
    \alpha&=\frac{(2-B^2a^2)^2}{\sqrt{4+B^4a^4+4B^2(m^2-a^2)}}\nonumber\\
    &\quad\times\sqrt{\frac{1-B^2a^2}{4+B^6a^6-3B^4a^4+4B^4m^2a^2}}\,,\\
    \Omega_{int}&=-B^2a\left[1+\frac{4B^2m^2}{(2-B^2a^2)^2}\right]\,,\\
    \Phi_{int}&=\frac{4B^3m^2a}{(2-B^2a^2)^2\sqrt{1-B^2a^2}}\,.
\end{align}
These parameters also exhibit the expected limits. In the absence of rotation ($a=0$), they reduce to the Schwarzschild-Bertotti-Robinson case,
\begin{align}
    \alpha=\frac{1}{\sqrt{1+B^2m^2}}\,,\qquad\Omega_{int}=0\,,\qquad\Phi_{int}=0\,.
\end{align}
On the other hand, when the external magnetic field vanishes ($B=0$), all parameters return to the Kerr values,
\begin{align}
    \alpha=1\,,\qquad\Omega_{int}=0\,,\qquad\Phi_{int}=0\,.
\end{align}

Furthermore, using the quantities obtained above, we introduce the following redefined thermodynamic potentials,
\begin{align}
    T=\alpha T_H\,,\,\quad\Omega=\alpha(\Omega_H-\Omega_{int})\,,\,\quad\Phi=\alpha(\Phi_H-\Phi_{int})\,.
\end{align}
These quantities are found to coincide with the thermodynamic potentials derived directly from the Christodoulou-Ruffini mass relation~\cite{Caldarelli:1999xj},
\begin{align}
    T&=\frac{\partial M}{\partial S}=\frac{1}{8\pi M}\left[1-\frac{\pi^2}{S^2}\left(4J^2+Q^4\right)\right]\,,\\
    \Omega&=\frac{\partial M}{\partial J}=\frac{\pi J}{MS}\,,\\
    \Phi&=\frac{\partial M}{\partial Q}=\frac{Q}{2MS}\left(S+\pi Q^2\right)\,.
\end{align}
With these identifications, the variation of the mass takes the standard form of the first law of black-hole thermodynamics (see Appendix~\ref{1st_law}),
\begin{align}
    \delta M=T\delta S+\Omega\delta J+\Phi\delta Q\,,
\end{align}
and consequently the Smarr formula,
\begin{align}
    M=2TS+2\Omega J+\Phi Q\,.
\end{align}

\section{Conclusions and discussions}

In this work, we have studied the thermodynamic properties of the Kerr-Bertotti-Robinson black hole. Using covariant phase space methods, we showed how the conserved angular momentum $J$ and electric charge $Q$ can be calculated. Owing to the presence of an asymptotically uniform external electromagnetic field, the standard formalism does not yield a well-defined conserved mass. To consistently address this issue, we have adopted the Christodoulou-Ruffini mass relation to resolve the ambiguity in the definition of the conserved mass. Based on this definition, we have determined the corresponding generator and constructed the redefined thermodynamic potentials. As a result, the thermodynamic quantities naturally satisfy the standard form of the first law of black-hole thermodynamics and the associated Smarr formula. This demonstrates that a consistent thermodynamic description can be established even in the presence of nontrivial asymptotic structures.

An interesting feature of our results is that, after identifying the conserved thermodynamic mass, neither the first law nor the Smarr formula contains an additional $\mu \delta B$ or $\mu B$ term, since all explicit dependence on the external magnetic field is absorbed into the thermodynamic quantities. This is consistent with previous analyses of the Kerr-Newman-Melvin black hole.


Finally,  the validity of Christodoulou-Ruffini mass relation is an assumption that need not be valid for Kerr-Bertotti-Robinson black hole \textit{a priori}, and whether the proposed thermodynamics will agree with thermodynamics derived from other prescriptions remains to be seen. It remains an interesting open question whether the same mass can be obtained from alternative approaches, such as the conformal method. We leave this for future work.

\begin{acknowledgments}
We thank Yu-Sen An, Marco Astorino, Shan-Ming Ruan, Anna Tokareva, and Jun-Kun Zhao for helpful discussions.
This work is supported by the National Key Research and Development Program of China Grants No. 2021YFC2203004, No. 2021YFA0718304, and No. 2020YFC2201501, the National Natural Science Foundation of China Grants No. 12422502, No. 12547110, No. 12588101, No. 12235019, and No. 12447101.
\end{acknowledgments}

\appendix

\section{First law of Kerr-Bertotti-Robinson black hole}\label{1st_law}

In this Appendix, we briefly show how the variation of the conserved mass can lead to the form of the first law.

According to Refs.~\cite{Compere:2006my, Astorino:2016hls}, the surface charge associated with the Killing generator of the black-hole horizon is independent of the choice of the spacelike integration surface. As a result,
\begin{align}
    \int_{H}k_{(\chi,0)}=\int_{\partial C}k_{(\chi,0)}\,.
\end{align}
where the left-hand side is evaluated on the horizon while the right-hand side is evaluated on a spacelike surface extending to infinity.

Evaluating the charge on the horizon gives the standard result~\cite{Bardeen:1973gs, Iyer:1994ys},
\begin{align}
    \int_{H}k_{(\chi,0)}=T_H\delta S+\Phi_H\delta Q\,.
\end{align}

To relate this expression to the conserved quantities defined at infinity, we decompose the gauge parameter linearly as
\begin{align}
\alpha(\chi,0)=\alpha(\partial_t+\Omega_{int}\partial_\varphi,\Phi_{int})+\alpha\left((\Omega_{H}-\Omega_{int})\partial_\varphi,-\Phi_{int}\right)\,.
\end{align}

Since the mapping between the surface charge and the gauge parameter is linear, the corresponding charges can be evaluated separately on the spacelike surfaces at the horizon and at infinity~\cite{Astorino:2016hls},
\begin{align}
    \int_H k_{(\alpha\xi,0)}&=\int_{\partial C}k_{\alpha(\partial_t+\Omega_{int}\partial_\varphi,\Phi_{int})}\nonumber\\
    &+\int_{\partial C} k_{\alpha((\Omega_H-\Omega_{int})\partial_\varphi,-\Phi_{int})}\,.
\end{align}

Using the definitions of the conserved charges
\begin{align}
    \delta Q&=\int_{\partial C}k_{(0,-1)}\,,\\
    \delta J&=\int_{\partial C}k_{(-\partial_{\varphi},0)}\,,\\
    \delta M&=\int_{\partial C}k_{\alpha(\partial_t+\Omega_{int}\partial_\varphi,\Phi_{int})}\,,
\end{align}
one finds
\begin{align}
    \delta M&=\alpha\Big(T_H\delta S+(\Omega_H-\Omega_{int})\delta J+(\Phi_H-\Phi_{int})\delta Q\Big)\,.
\end{align}

With the redefined thermodynamic parameters $T$, $\Omega$, and $\Phi$, the above first law can be written in the standard form,
\begin{align}
    \delta M=T\delta S+\Omega\delta J+\Phi\delta Q\,.
\end{align}

\bibliography{biblatex}

@article{Podolsky:2025tle,
    author = "Podolsky, Jiri and Ovcharenko, Hryhorii",
    title = "{Kerr Black Hole in a Uniform Bertotti-Robinson Magnetic Field: An Exact Solution}",
    eprint = "2507.05199",
    archivePrefix = "arXiv",
    primaryClass = "gr-qc",
    doi = "10.1103/rfgv-ybz5",
    journal = "Phys. Rev. Lett.",
    volume = "135",
    number = "18",
    pages = "181401",
    year = "2025"
}

@article{Ovcharenko:2025cpm,
    author = "Ovcharenko, Hryhorii and Podolsk{\'y}, Ji{\v{r}}{\'\i}",
    title = "{New class of rotating charged black holes with nonaligned electromagnetic field}",
    eprint = "2508.04850",
    archivePrefix = "arXiv",
    primaryClass = "gr-qc",
    doi = "10.1103/8wkz-th6v",
    journal = "Phys. Rev. D",
    volume = "112",
    number = "6",
    pages = "064076",
    year = "2025"
}

@article{Compere:2009dp,
    author = "Compere, Geoffrey and Murata, Keiju and Nishioka, Tatsuma",
    title = "{Central Charges in Extreme Black Hole/CFT Correspondence}",
    eprint = "0902.1001",
    archivePrefix = "arXiv",
    primaryClass = "hep-th",
    reportNumber = "KUNS-2187",
    doi = "10.1088/1126-6708/2009/05/077",
    journal = "JHEP",
    volume = "05",
    pages = "077",
    year = "2009"
}

@article{Barnich:2001jy,
    author = "Barnich, Glenn and Brandt, Friedemann",
    title = "{Covariant theory of asymptotic symmetries, conservation laws and central charges}",
    eprint = "hep-th/0111246",
    archivePrefix = "arXiv",
    reportNumber = "ULB-TH-01-19, MPI-MIS-94-2001",
    doi = "10.1016/S0550-3213(02)00251-1",
    journal = "Nucl. Phys. B",
    volume = "633",
    pages = "3--82",
    year = "2002"
}

@article{Barnich:2003xg,
    author = "Barnich, Glenn",
    title = "{Boundary charges in gauge theories: Using Stokes theorem in the bulk}",
    eprint = "hep-th/0301039",
    archivePrefix = "arXiv",
    reportNumber = "ULB-TH-03-01",
    doi = "10.1088/0264-9381/20/16/310",
    journal = "Class. Quant. Grav.",
    volume = "20",
    pages = "3685--3698",
    year = "2003"
}

@article{Iyer:1994ys,
    author = "Iyer, Vivek and Wald, Robert M.",
    title = "{Some properties of Noether charge and a proposal for dynamical black hole entropy}",
    eprint = "gr-qc/9403028",
    archivePrefix = "arXiv",
    doi = "10.1103/PhysRevD.50.846",
    journal = "Phys. Rev. D",
    volume = "50",
    pages = "846--864",
    year = "1994"
}

@inproceedings{Compere:2006my,
    author = "Compere, Geoffrey",
    title = "{An introduction to the mechanics of black holes}",
    booktitle = "{2nd Modave Summer School in Theoretical Physics}",
    eprint = "gr-qc/0611129",
    archivePrefix = "arXiv",
    reportNumber = "ULB-TH-06-29",
    month = "11",
    year = "2006"
}

@article{Astorino:2016ybm,
    author = "Astorino, Marco",
    title = "{Thermodynamics of Regular Accelerating Black Holes}",
    eprint = "1612.04387",
    archivePrefix = "arXiv",
    primaryClass = "gr-qc",
    reportNumber = "UAI-PHY-16-08",
    doi = "10.1103/PhysRevD.95.064007",
    journal = "Phys. Rev. D",
    volume = "95",
    number = "6",
    pages = "064007",
    year = "2017"
}

@article{Astorino:2025lih,
    author = "Astorino, Marco",
    title = "{Black holes in the external Bertotti-Robinson-Bonnor-Melvin electromagnetic field}",
    eprint = "2508.12908",
    archivePrefix = "arXiv",
    primaryClass = "gr-qc",
    reportNumber = "LIFT-11-4.25",
    doi = "10.1103/c5lw-53yd",
    journal = "Phys. Rev. D",
    volume = "112",
    number = "10",
    pages = "104077",
    year = "2025"
}

@article{Astorino:2016xiy,
    author = "Astorino, Marco",
    title = "{CFT Duals for Accelerating Black Holes}",
    eprint = "1605.06131",
    archivePrefix = "arXiv",
    primaryClass = "hep-th",
    reportNumber = "UAI-PHY-16-06",
    doi = "10.1016/j.physletb.2016.07.019",
    journal = "Phys. Lett. B",
    volume = "760",
    pages = "393--405",
    year = "2016"
}

@article{Astorino:2016hls,
    author = "Astorino, M. and Comp{\`e}re, G. and Oliveri, R. and Vandevoorde, N.",
    title = "{Mass of Kerr-Newman black holes in an external magnetic field}",
    eprint = "1602.08110",
    archivePrefix = "arXiv",
    primaryClass = "gr-qc",
    doi = "10.1103/PhysRevD.94.024019",
    journal = "Phys. Rev. D",
    volume = "94",
    number = "2",
    pages = "024019",
    year = "2016"
}

@article{Bardeen:1973gs,
    author = "Bardeen, James M. and Carter, B. and Hawking, S. W.",
    title = "{The Four laws of black hole mechanics}",
    doi = "10.1007/BF01645742",
    journal = "Commun. Math. Phys.",
    volume = "31",
    pages = "161--170",
    year = "1973"
}

@article{Siahaan:2025ngu,
    author = "Siahaan, Haryanto M.",
    title = "{Kerr{\textendash}Bertotti{\textendash}Robinson spacetime and the Kerr/CFT correspondence}",
    eprint = "2512.12533",
    archivePrefix = "arXiv",
    primaryClass = "gr-qc",
    doi = "10.1016/j.nuclphysb.2026.117514",
    journal = "Nucl. Phys. B",
    volume = "1028",
    pages = "117514",
    year = "2026"
}

@article{Bekenstein:1973ur,
    author = "Bekenstein, Jacob D.",
    title = "{Black holes and entropy}",
    doi = "10.1103/PhysRevD.7.2333",
    journal = "Phys. Rev. D",
    volume = "7",
    pages = "2333--2346",
    year = "1973"
}

@article{EventHorizonTelescope:2019dse,
    author = "Akiyama, Kazunori and others",
    collaboration = "Event Horizon Telescope",
    title = "{First M87 Event Horizon Telescope Results. I. The Shadow of the Supermassive Black Hole}",
    eprint = "1906.11238",
    archivePrefix = "arXiv",
    primaryClass = "astro-ph.GA",
    doi = "10.3847/2041-8213/ab0ec7",
    journal = "Astrophys. J. Lett.",
    volume = "875",
    pages = "L1",
    year = "2019"
}

@article{LIGOScientific:2016aoc,
    author = "Abbott, B. P. and others",
    collaboration = "LIGO Scientific, Virgo",
    title = "{Observation of Gravitational Waves from a Binary Black Hole Merger}",
    eprint = "1602.03837",
    archivePrefix = "arXiv",
    primaryClass = "gr-qc",
    reportNumber = "LIGO-P150914",
    doi = "10.1103/PhysRevLett.116.061102",
    journal = "Phys. Rev. Lett.",
    volume = "116",
    number = "6",
    pages = "061102",
    year = "2016"
}

@article{Richstone:1998ky,
    author = "Richstone, D. and others",
    title = "{Supermassive black holes and the evolution of galaxies}",
    eprint = "astro-ph/9810378",
    archivePrefix = "arXiv",
    journal = "Nature",
    volume = "395",
    pages = "A14--A19",
    year = "1998"
}

@article{Kormendy:2013dxa,
    author = "Kormendy, John and Ho, Luis C.",
    title = "{Coevolution (Or Not) of Supermassive Black Holes and Host Galaxies}",
    eprint = "1304.7762",
    archivePrefix = "arXiv",
    primaryClass = "astro-ph.CO",
    doi = "10.1146/annurev-astro-082708-101811",
    journal = "Ann. Rev. Astron. Astrophys.",
    volume = "51",
    pages = "511--653",
    year = "2013"
}

@article{Hu:2023oiu,
    author = "Hu, Li and Cai, Rong-Gen and Wang, Shao-Jiang",
    title = "{Distinctive GWBs from eccentric inspiraling SMBH binaries with a DM spike}",
    eprint = "2312.14041",
    archivePrefix = "arXiv",
    primaryClass = "gr-qc",
    doi = "10.1088/1475-7516/2025/02/067",
    journal = "JCAP",
    volume = "02",
    pages = "067",
    year = "2025"
}

@article{Hu:2024hzu,
    author = "Hu, Li and Cai, Rong-Gen and Wang, Shao-Jiang",
    title = "{Dynamical friction can flip the hierarchical three-body system}",
    eprint = "2411.14047",
    archivePrefix = "arXiv",
    primaryClass = "gr-qc",
    doi = "10.1088/1475-7516/2025/08/010",
    journal = "JCAP",
    volume = "08",
    pages = "010",
    year = "2025"
}

@article{Vicente:2025gsg,
    author = "Vicente, Rodrigo and Karydas, Theophanes K. and Bertone, Gianfranco",
    title = "{Fully Relativistic Treatment of Extreme Mass-Ratio Inspirals in Collisionless Environments}",
    eprint = "2505.09715",
    archivePrefix = "arXiv",
    primaryClass = "gr-qc",
    doi = "10.1103/s4wh-x6c4",
    journal = "Phys. Rev. Lett.",
    volume = "135",
    number = "21",
    pages = "211401",
    year = "2025"
}

@article{You:2023dax,
    author = "You, Bei and others",
    title = "{Observations of a black hole x-ray binary indicate formation of a magnetically arrested disk}",
    eprint = "2309.00200",
    archivePrefix = "arXiv",
    primaryClass = "astro-ph.HE",
    doi = "10.1126/science.abo4504",
    journal = "Science",
    volume = "381",
    number = "6661",
    pages = "abo4504",
    year = "2023"
}

@article{Eatough:2013nva,
    author = "Eatough, R. P. and others",
    title = "{A strong magnetic field around the supermassive black hole at the centre of the Galaxy}",
    eprint = "1308.3147",
    archivePrefix = "arXiv",
    primaryClass = "astro-ph.GA",
    doi = "10.1038/nature12499",
    journal = "Nature",
    volume = "501",
    pages = "391--394",
    year = "2013"
}

@article{Ernst:1976bsr,
    author = "Ernst, Frederick J. and Wild, Walter J.",
    title = "{Kerr black holes in a magnetic universe}",
    doi = "10.1063/1.522875",
    journal = "J. Math. Phys.",
    volume = "17",
    number = "2",
    pages = "182",
    year = "1976"
}

@article{Ernst:1976mzr,
    author = "Ernst, Frederick J.",
    title = "{Black holes in a magnetic universe}",
    doi = "10.1063/1.522781",
    journal = "J. Math. Phys.",
    volume = "17",
    number = "1",
    pages = "54--56",
    year = "1976"
}

@article{Pravda:2005uv,
    author = "Pravda, Vojtech and Zaslavskii, O. B.",
    title = "{Curvature tensors on distorted Killing horizons and their algebraic classification}",
    eprint = "gr-qc/0510095",
    archivePrefix = "arXiv",
    doi = "10.1088/0264-9381/22/23/009",
    journal = "Class. Quant. Grav.",
    volume = "22",
    pages = "5053--5072",
    year = "2005"
}

@article{Gibbons:2013yq,
    author = "Gibbons, G. W. and Mujtaba, A. H. and Pope, C. N.",
    title = "{Ergoregions in Magnetised Black Hole Spacetimes}",
    eprint = "1301.3927",
    archivePrefix = "arXiv",
    primaryClass = "gr-qc",
    reportNumber = "DAMTP-2013-5, MIFPA-13-02",
    doi = "10.1088/0264-9381/30/12/125008",
    journal = "Class. Quant. Grav.",
    volume = "30",
    number = "12",
    pages = "125008",
    year = "2013"
}

@article{Zeng:2025olq,
    author = "Zeng, Xiao-Xiong and Wang, Ke",
    title = "{Energy extraction from the Kerr-Bertotti-Robinson black hole via magnetic reconnection in a circular and a plunging plasma}",
    eprint = "2507.21777",
    archivePrefix = "arXiv",
    primaryClass = "gr-qc",
    doi = "10.1103/vc96-snjm",
    journal = "Phys. Rev. D",
    volume = "112",
    number = "6",
    pages = "064032",
    year = "2025"
}

@article{Wang:2025vsx,
    author = "Wang, Xinyu and Hou, Yehui and Wan, Xi and Guo, Minyong and Chen, Bin",
    title = "{Geodesics and shadows in the Kerr-Bertotti-Robinson black hole spacetime}",
    eprint = "2507.22494",
    archivePrefix = "arXiv",
    primaryClass = "gr-qc",
    doi = "10.1088/1475-7516/2026/02/050",
    journal = "JCAP",
    volume = "02",
    pages = "050",
    year = "2026"
}

@article{Ali:2025beh,
    author = "Ali, Heena and Ghosh, Sushant G.",
    title = "{Parameter estimation of Kerr-Bertotti-Robinson black holes using their shadows}",
    eprint = "2508.15862",
    archivePrefix = "arXiv",
    primaryClass = "gr-qc",
    doi = "10.1088/1475-7516/2026/01/018",
    journal = "JCAP",
    volume = "01",
    pages = "018",
    year = "2026"
}

@article{Li:2025rtf,
    author = "Li, Xiang-Qian and Yan, Hao-Peng and Yue, Xiao-Jun",
    title = "{Gravitational-wave imprints of Kerr{\textendash}Bertotti{\textendash}Robinson black holes: frequency blue-shift and waveform dephasing}",
    eprint = "2512.02921",
    archivePrefix = "arXiv",
    primaryClass = "gr-qc",
    doi = "10.1140/epjc/s10052-026-15441-5",
    journal = "Eur. Phys. J. C",
    volume = "86",
    number = "2",
    pages = "176",
    year = "2026"
}

@article{Xamidov:2026kqs,
    author = "Xamidov, Tursunali and Shaymatov, Sanjar and Wu, Qiang and Zhu, Tao",
    title = "{Gravitational wave signatures from periodic orbits around a Schwarzschild-Bertotti-Robinson black hole}",
    eprint = "2602.09453",
    archivePrefix = "arXiv",
    primaryClass = "gr-qc",
    month = "2",
    year = "2026"
}

@article{Mirkhaydarov:2026fyn,
    author = "Mirkhaydarov, Mirjavoxir and Xamidov, Tursunali and Sheoran, Pankaj and Shaymatov, Sanjar and Nandan, Hemwati",
    title = "{Non-Monotonic Enhancement of the Magnetic Penrose Process in Kerr-Bertotti-Robinson Spacetime and its Implication for Electron Acceleration}",
    eprint = "2601.09919",
    archivePrefix = "arXiv",
    primaryClass = "gr-qc",
    month = "1",
    year = "2026"
}

@article{Yuan:2026knu,
    author = "Yuan, Xulong and Zhang, Xiangdong",
    title = "{External magnetic field influence on massive binary black hole inspiral gravitational waves and its similarity with environmental effects}",
    eprint = "2603.05084",
    archivePrefix = "arXiv",
    primaryClass = "gr-qc",
    month = "3",
    year = "2026"
}

@article{Christodoulou:1971pcn,
    author = "Christodoulou, D. and Ruffini, R.",
    title = "{Reversible transformations of a charged black hole}",
    doi = "10.1103/PhysRevD.4.3552",
    journal = "Phys. Rev. D",
    volume = "4",
    pages = "3552--3555",
    year = "1971"
}

@article{Hu:2025bbc,
    author = "Hu, Li and Cai, Rong-Gen and Wang, Shao-Jiang",
    title = "{Third law of repetitive electric Penrose processes}",
    eprint = "2510.26866",
    archivePrefix = "arXiv",
    primaryClass = "gr-qc",
    doi = "10.1103/vlgc-y3qg",
    journal = "Phys. Rev. D",
    volume = "113",
    number = "6",
    pages = "L061501",
    year = "2026"
}

@article{Gao:2023luj,
    author = "Gao, Yunjiao and Di, Zhenbo and Gao, Sijie",
    title = "{General mass formulas for charged Kerr-AdS black holes}",
    eprint = "2304.10290",
    archivePrefix = "arXiv",
    primaryClass = "gr-qc",
    doi = "10.1088/1402-4896/ad6fff",
    journal = "Phys. Scripta",
    volume = "99",
    number = "9",
    pages = "095022",
    year = "2024"
}

@article{Gibbons:2013dna,
    author = "Gibbons, G. W. and Pang, Yi and Pope, C. N.",
    title = "{Thermodynamics of magnetized Kerr-Newman black holes}",
    eprint = "1310.3286",
    archivePrefix = "arXiv",
    primaryClass = "hep-th",
    reportNumber = "MIFPA-13-28",
    doi = "10.1103/PhysRevD.89.044029",
    journal = "Phys. Rev. D",
    volume = "89",
    number = "4",
    pages = "044029",
    year = "2014"
}

@article{Caldarelli:1999xj,
    author = "Caldarelli, Marco M. and Cognola, Guido and Klemm, Dietmar",
    title = "{Thermodynamics of Kerr-Newman-AdS black holes and conformal field theories}",
    eprint = "hep-th/9908022",
    archivePrefix = "arXiv",
    reportNumber = "UTF-434",
    doi = "10.1088/0264-9381/17/2/310",
    journal = "Class. Quant. Grav.",
    volume = "17",
    pages = "399--420",
    year = "2000"
}

@article{Lyu:2025lue,
    author = "Lyu, Zhen-Hong and Cai, Rong-Gen and Guo, Zong-Kuan and He, Jian-Feng and Liu, Jing",
    title = "{Ring formation from black hole superradiance through repeated particle production on bound orbits}",
    eprint = "2507.03490",
    archivePrefix = "arXiv",
    primaryClass = "gr-qc",
    doi = "10.1103/3r41-xyj3",
    journal = "Phys. Rev. D",
    volume = "112",
    number = "10",
    pages = "104066",
    year = "2025"
}

@article{Zeng:2025tji,
    author = "Zeng, Xiao-Xiong and Yang, Chen-Yu and Yu, Hao",
    title = "{Optical characteristics of the Kerr{\textendash}Bertotti{\textendash}Robinson black hole}",
    eprint = "2508.03020",
    archivePrefix = "arXiv",
    primaryClass = "gr-qc",
    doi = "10.1140/epjc/s10052-025-14989-y",
    journal = "Eur. Phys. J. C",
    volume = "85",
    number = "11",
    pages = "1242",
    year = "2025"
}

@article{Wang:2025bjf,
    author = "Wang, Tower",
    title = "{Innermost stable circular orbit of Kerr-Bertotti-Robinson black holes and inspirals from it: Exact solutions}",
    eprint = "2508.04684",
    archivePrefix = "arXiv",
    primaryClass = "gr-qc",
    month = "8",
    year = "2025"
}

@article{Vachher:2025jsq,
    author = "Vachher, Amnish and Kumar, Arun and Ghosh, Sushant G.",
    title = "{The influence of uniform magnetic fields on strong field gravitational lensing by Kerr black holes}",
    eprint = "2508.21100",
    archivePrefix = "arXiv",
    primaryClass = "gr-qc",
    doi = "10.1088/1475-7516/2025/11/021",
    journal = "JCAP",
    volume = "11",
    pages = "021",
    year = "2025"
}

@article{Zhang:2025ole,
    author = "Zhang, Yu-Kun and Wei, Shao-Wen",
    title = "{Effects of magnetic fields on spinning test particles orbiting Kerr-Bertotti-Robinson black holes}",
    eprint = "2510.07914",
    archivePrefix = "arXiv",
    primaryClass = "gr-qc",
    doi = "10.1103/dmh3-ht32",
    journal = "Phys. Rev. D",
    volume = "113",
    number = "10",
    pages = "104024",
    year = "2026"
}

@article{Liu:2025wwq,
    author = "Liu, Wentao and Liu, Yang and Wu, Di and Liu, Yu-Xiao",
    title = "{Universal framework for horizon-scale tests of gravity with black hole shadows}",
    eprint = "2511.06017",
    archivePrefix = "arXiv",
    primaryClass = "gr-qc",
    doi = "10.1103/66x9-zj16",
    journal = "Phys. Rev. D",
    volume = "114",
    number = "2",
    pages = "L021503",
    year = "2026"
}

@article{Ahmed:2025ril,
    author = "Ahmed, Faizuddin and Sakall{\i}, {\.I}zzet and Al-Badawi, Ahmad",
    title = "{Kerr-Bertotti-Robinson Black Holes Surrounded by a Cloud of Strings}",
    eprint = "2511.11792",
    archivePrefix = "arXiv",
    primaryClass = "gr-qc",
    month = "11",
    year = "2025"
}

@article{Ortaggio:2025sip,
    author = "Ortaggio, Marcello",
    title = "{Einstein-Maxwell fields as solutions of Einstein gravity coupled to conformally invariant nonlinear electrodynamics}",
    eprint = "2511.13665",
    archivePrefix = "arXiv",
    primaryClass = "gr-qc",
    doi = "10.1103/nwcq-n4hg",
    journal = "Phys. Rev. D",
    volume = "113",
    number = "8",
    pages = "L081503",
    year = "2026"
}

@article{Gray:2025lwy,
    author = "Gray, Finnian and Kubiznak, David and Ovcharenko, Hryhorii and Podolsky, Jiri",
    title = "{Hidden symmetries and separability structures of Ovcharenko-Podolsk{\'y} and conformal-to-Carter spacetimes}",
    eprint = "2511.21538",
    archivePrefix = "arXiv",
    primaryClass = "gr-qc",
    doi = "10.1103/8832-htpg",
    journal = "Phys. Rev. D",
    volume = "113",
    number = "4",
    pages = "044050",
    year = "2026"
}

@article{Andersson:2025bhq,
    author = "Andersson, Lars and Gray, Finnian and Oancea, Marius A.",
    title = "{Conserved quantities and integrability for massless spinning particles in general relativity}",
    eprint = "2512.07677",
    archivePrefix = "arXiv",
    primaryClass = "gr-qc",
    doi = "10.1103/85dv-n8tr",
    journal = "Phys. Rev. D",
    volume = "113",
    number = "6",
    pages = "064061",
    year = "2026"
}

@article{Astorino:2026okd,
    author = "Astorino, Marco",
    title = "{Static hairy black hole in 4D general relativity}",
    eprint = "2601.16254",
    archivePrefix = "arXiv",
    primaryClass = "gr-qc",
    reportNumber = "LIFT-12-5.25",
    doi = "10.1103/yz86-wc3g",
    journal = "Phys. Rev. D",
    volume = "113",
    number = "2",
    pages = "024047",
    year = "2026"
}

@article{Barrientos:2026kdl,
    author = {Barrientos, Jos{\'e} and Canfora, Fabrizio and Cisterna, Adolfo and M{\"u}ller, Keanu and Neira, Anibal},
    title = "{Melvin{\textendash}Bonnor and Bertotti{\textendash}Robinson spacetimes with baryonic charge}",
    eprint = "2601.19858",
    archivePrefix = "arXiv",
    primaryClass = "gr-qc",
    doi = "10.1140/epjc/s10052-026-15937-0",
    journal = "Eur. Phys. J. C",
    volume = "86",
    number = "6",
    pages = "673",
    year = "2026"
}

@article{Wang:2026czl,
    author = "Wang, Chao-Hui and Meng, Xiang-Cheng and Wei, Shao-Wen",
    title = "{Magnetic field effects on spherical orbit in Kerr-Bertotti-Robinson spacetime: constraints from jet precession of M87*}",
    eprint = "2602.03161",
    archivePrefix = "arXiv",
    primaryClass = "gr-qc",
    month = "2",
    year = "2026"
}

@article{Mustafa:2026gly,
    author = "Mustafa, G. and Donmez, Orhan and Gogoi, Dhruba Jyoti and Ghosh, Sushant G. and Hussain, Ibrar and Yuan, Chengxun",
    title = "{Dynamics, ringdown, and accretion-driven multiple quasi-periodic oscillations of Kerr{\textendash}Bertotti{\textendash}Robinson black holes}",
    eprint = "2602.08911",
    archivePrefix = "arXiv",
    primaryClass = "gr-qc",
    doi = "10.1088/1475-7516/2026/07/065",
    journal = "JCAP",
    volume = "07",
    pages = "065",
    year = "2026"
}

@article{Ovcharenko:2026byw,
    author = "Ovcharenko, Hryhorii and Podolsky, Jiri",
    title = {{Static black holes in an external uniform electromagnetic field: Reissner-Nordstr{\"o}m black holes accelerating in Bertotti-Robinson spacetime}},
    eprint = "2602.15462",
    archivePrefix = "arXiv",
    primaryClass = "gr-qc",
    doi = "10.1103/824d-8zbj",
    journal = "Phys. Rev. D",
    volume = "114",
    number = "2",
    pages = "024068",
    year = "2026"
}

@article{Barrientos:2026shy,
    author = {Barrientos, Jos{\'e} and Cisterna, Adolfo and D{\'\i}az, Amaro and M{\"u}ller, Keanu},
    title = "{From Bertotti--Robinson to Vacuum: New Exact Solutions in General Relativity via Harrison and Inversion Symmetries}",
    eprint = "2602.17581",
    archivePrefix = "arXiv",
    primaryClass = "gr-qc",
    month = "2",
    year = "2026"
}

@article{Siahaan:2026tuf,
    author = "Siahaan, Haryanto M.",
    title = "{Meissner Effect in Kerr--Bertotti--Robinson Spacetime}",
    eprint = "2603.00653",
    archivePrefix = "arXiv",
    primaryClass = "gr-qc",
    month = "2",
    year = "2026"
}

@article{Al-Badawi:2026whl,
    author = "Al-Badawi, Ahmad and Ahmed, Faizuddin and Silva, Edilberto O.",
    title = "{Accelerating Bertotti-Robinson Black Holes in a Uniform Magnetic Field}",
    eprint = "2603.03494",
    archivePrefix = "arXiv",
    primaryClass = "gr-qc",
    month = "3",
    year = "2026"
}

@article{Lu:2026kcm,
    author = "Lu, Junjie and Wu, Xin",
    title = "{Third type of spacetime with the coexistence of integrability and non-integrability}",
    eprint = "2603.12674",
    archivePrefix = "arXiv",
    primaryClass = "gr-qc",
    doi = "10.1140/epjc/s10052-026-15482-w",
    journal = "Eur. Phys. J. C",
    volume = "86",
    number = "3",
    pages = "256",
    year = "2026"
}

\end{document}